# COVID-19 Detection in Chest X-Ray Images using a New Channel Boosted CNN

Saddam Hussain Khan[1,2], Anabia Sohail[1,2], and Asifullah Khan*[1,2,3]
[1]Pattern Recognition Lab, Department of Computer & Information Sciences, Pakistan Institute of Engineering & Applied Sciences, Nilore, Islamabad 45650, Pakistan,
[2]PIEAS Artificial Intelligence Center (PAIC), Pakistan Institute of Engineering & Applied Sciences, Nilore, Islamabad 45650, Pakistan
[3]Center for Mathematical Sciences, Pakistan Institute of Engineering & Applied Sciences, Nilore, Islamabad 45650, Pakistan
asif@pieas.edu.pk

## Abstract

COVID-19 is a highly contagious respiratory infection that has affected a large population across the world and continues with its devastating consequences. It is imperative to detect COVID-19 at the earliest to limit the span of infection. In this work, a new classification technique "CB-STM-RENet" based on deep Convolutional Neural Network (CNN) and Channel Boosting is proposed for the screening of COVID-19 in chest X-Rays. In this connection, to learn the COVID-19 specific radiographic patterns, a new convolution block based on split-transform-merge (STM) is developed. This new block systematically incorporates region and edge-based operations at each branch to capture the diverse set of features at various levels, especially those related to region homogeneity, textural variations, and boundaries of the infected region. The learning and discrimination capability of the proposed CNN architecture is enhanced by exploiting the Channel Boosting idea that concatenates the auxiliary channels along with the original channels. The auxiliary channels are generated from the pre-trained CNNs using Transfer Learning. The effectiveness of the proposed technique "CB-STM-RENet" is evaluated on three different datasets of chest X-Rays namely CoV-Healthy-6k, CoV-NonCoV-10k, and CoV-NonCoV-15k. The performance comparison of the proposed CB-STM-RENet with the existing techniques exhibits high performance both in discriminating COVID-19 chest infections from Healthy, as well as, other types of chest infections. CB-STM-RENet provides the highest performance on all these three datasets; especially on the stringent CoV-NonCoV-15k dataset. The good detection rate (97%), and high precision (93%) of the proposed technique suggest that it can be adapted for the diagnosis of COVID-19 infected patients. The test code is available at https://github.com/PRLAB21/COVID-19-Detection-System-using-Chest-X-Ray-Images.

**Keywords:** Coronavirus, COVID-19, SARS-CoV-2, Chest X-ray, Channel boosting, Split-transform-Merge, Convolutional Neural Network, and Transfer Learning.

# 1. Introduction

Coronavirus disease 2019 (COVID-19) is a severe and continuing pandemic, which broke out in December 2019 and has now affected the whole world. This new pathogenic viral infection is caused by a new virus from the family of coronavirus (CoV) and named as SARS-CoV-2. COVID-19 is highly contagious, which quickly transmits from one individual to another, even before the onset of clinical symptoms [1], [2]. COVID-19 causes a respiratory illness that can be asymptomatic, or its clinical manifestation can span across fever, cough, myalgia, respiratory impairment, pneumonia, acute respiratory distress and even death in severe cases [3], [4].

The lack of a standard vaccine and no approved treatment by the Food and Drug Authority necessitates the early detection of COVID-19 both for proper care of patients and to control infection spread. The standard approach approved by World Health Organization for virus antigen detection is Polymerase Chain Reaction (PCR); however, it suffers from False-negative rate depending upon viral load and sampling strategy (30–70% True-positive rate) [5], [6].

Radiological imaging (X-Ray, CT) is used as an assisted screening tool to counter the False-negative rate of PCR in symptomatic patients. It acts as a first-line diagnostic measure for patients suspected with COVID-19 and suffering from a chest infection [7]. In addition to diagnostic importance, X-Ray and CT images are used for severity assessment and patients' follow-up [8], [9]. Chest imaging manifests radiological patterns specific to COVID-19. These patterns commonly include multifocal and ground-glass opacities and multi-lobular involvement. In patients with severe COVID-19 pneumonia, lung density increases, and its characteristic marks become incomprehensible because of consolidation [10], [11]. X-Ray imaging, as compared to CT imaging, is a quick and easy method that is widely available at low cost. The facility of X-Ray imaging is commonly available in hospitals, and the availability of its portable devices make it easy to perform X-Rays imaging in deprived areas, field hospitals and intensive care units [12].

The visual assessments of radiographic images of COVID-19 patients require trained radiologists. In an ongoing pandemic, the high prevalence of COVID-19 cases in addition to other pulmonary disorders and a limited number of experts is a considerable burden on radiologists. The inevitable importance of a timely diagnosis stresses the need for the development of automated assistance tool that can facilitate radiologists in the initial screening.

Deep learning (DL) models are a powerful tool for the analysis of medical images. DL models have been successfully used for the analysis of thoracic radiologic images for diagnosis of pneumonia, respiratory distress, tuberculosis, and segmentation of infected regions [13]–[17].

In this work, a deep Channel Boosted (CB) CNN based new classification technique "CB-STM-RENet" is proposed for the automatic detection of COVID-19 in chest X-Rays. CB-STM-RENet is a new CNN architecture and is able to discriminate both COVID-19 chest infections from healthy, as well as other types of chest infections. The proposed technique exploits the systematic usage of the region and edge-based (RE) feature extraction in our newly proposed convolutional block. Additionally, to enhance the classification performance, the concept of Channel Boosting is exploited. In this regard, for the creation of the boosted channels, Transfer Learning (TL) based fine-tuned pre-trained CNN models are used as auxiliary learners. This combination of the original and auxiliary channels boosts the chest infection discrimination capability of the proposed CB-STM-RENet. The significant contributions of this research are:

1. A novel CNN block based on the concept of Split-Transform-Merge (STM) is developed that systematically exploits the concept of RE-based (RE) feature extraction in each block of the proposed STM-RENet.
2. The systematic use of the RE-based operations at each branch of the STM block captures the diverse set of features at various levels, especially those related to region homogeneity, textural variations, and boundaries of the infected region.
3. The idea of Channel Boosting is exploited using TL in addition to the new STM block to reduce the False negatives and thus achieved a significant improvement in STM-RENet performance.
4. The final proposed classification technique "CB-STM-RENet" thus exploits the effectiveness of a diverse set of RE-based features, Channel Boosting, and TL.
5. We have also assembled three different datasets from the publicly available chest X-Ray images and named them as CoV-Healthy-6k, CoV-NonCoV-10k, and CoV-NonCoV-15k, respectively.

The remainder of this paper is structured as follows. Related work is presented in section 2. Dataset description is given in section 3. The detailed framework of the proposed deep CB-CNN for COVID-19 detection is explained in section 4. Section 5 provides the experimental setup. Section 6 discusses the results and comparative studies. Finally, section 7 concludes the paper.

## 2. Related Work

DL models have shown impressive performance in the medical field. Therefore, DL is primarily employed in the detection of COVID-19 infection in diverse ways. Several researchers have employed CNN to speed up the analysis of COVID-19 infected images [18]–[21]. Initially, COVID-19 labelled datasets were small in size and generally not suitable for practical implementation. Therefore, the existing pre-trained CNN architectures such as AlexNet, VGG, ResNet, Dense Net, Inception etc. have been employed on COVID-19 classification challenge. These architectures have been fine-tuned on problem-specific COVID dataset using TL and achieved optimal results. However, because of the non-availability of the consolidated data repository, these models have been evaluated on various small size datasets gathered from GitHub and Open-I respiratory, etc.

In one of the early works, a pre-trained ResNet-50 model has been fine-tuned using TL on small X-ray chest dataset and achieved 98% accuracy [16]. Similarly, a pre-trained ResNet-101 CNN architecture has been used to detect abnormality in chest X-ray images using different dataset and reported sensitivity (0.77), and accuracy (71.9%), respectively [22]. Moreover, a pre-trained inception network has been employed for the prediction of COVID-19 and reported accuracy (89.5%) [23]. The model has been employed on the multi-class problem like Healthy, COVID-19 non-affected patients, and COVID-19 affected pneumonia patients.

Similarly, nineteen layers of deep CNN architecture has been developed based on the idea of ResNet, named as COVID-Net and employed on the same dataset [24]. COVID-Net model showed good accuracy (92%) but at a low detection rate (sensitivity) (87%). Similarly, COVID-CAPS designed based on the concept of Capsule Net and achieved accuracy (98%), sensitivity (0.80), and AUC (0.97) [25]. An existing deep CNN Darknet network has been proposed for an accurate diagnosis of COVID infection on the complex dataset. In this work, darknet performed both binary (COVID-19 vs. Non-COVID-19) and multi-class (COVID-19, Pneumonia, and Healthy) discrimination. Darknet achieved a detection accuracy (98%) and (87%) for binary and multi-class, respectively [26].

In the most recent studies, a framework of four pre-trained existing CNN networks has been employed for identifying the presence of COVID-19 in X-ray images. These models (ResNet18, ResNet50, Squeeze Net, and DenseNet121) are fine-tuned on COVID-Xray-5k dataset. On average, these models obtained approximately a detection rate of (98%) [27]. A pre-trained CNN model like ResNet-50 has also been used for deep feature extraction and ML classification (SVM). This model is fine-tuned on small COVID-19 dataset using TL and reported an accuracy of 95% [28]. Similarly, a pre-trained ResNet-152 has been used for deep feature extraction in combination with Random Forest and XGBoost classifiers, achieving accuracy (97.3%) and (97.7%), respectively [29].

## 3. Dataset Details

COVID-19 is a new disease and to the best of our knowledge, up till now, no consolidated data is available. Consequently, we collected radiologists' authenticated X-Rays images from different publically accessible data repositories. The details of the datasets are mentioned in this section. The examples of X-Rays images from the assembled dataset are illustrated in Figure 1.

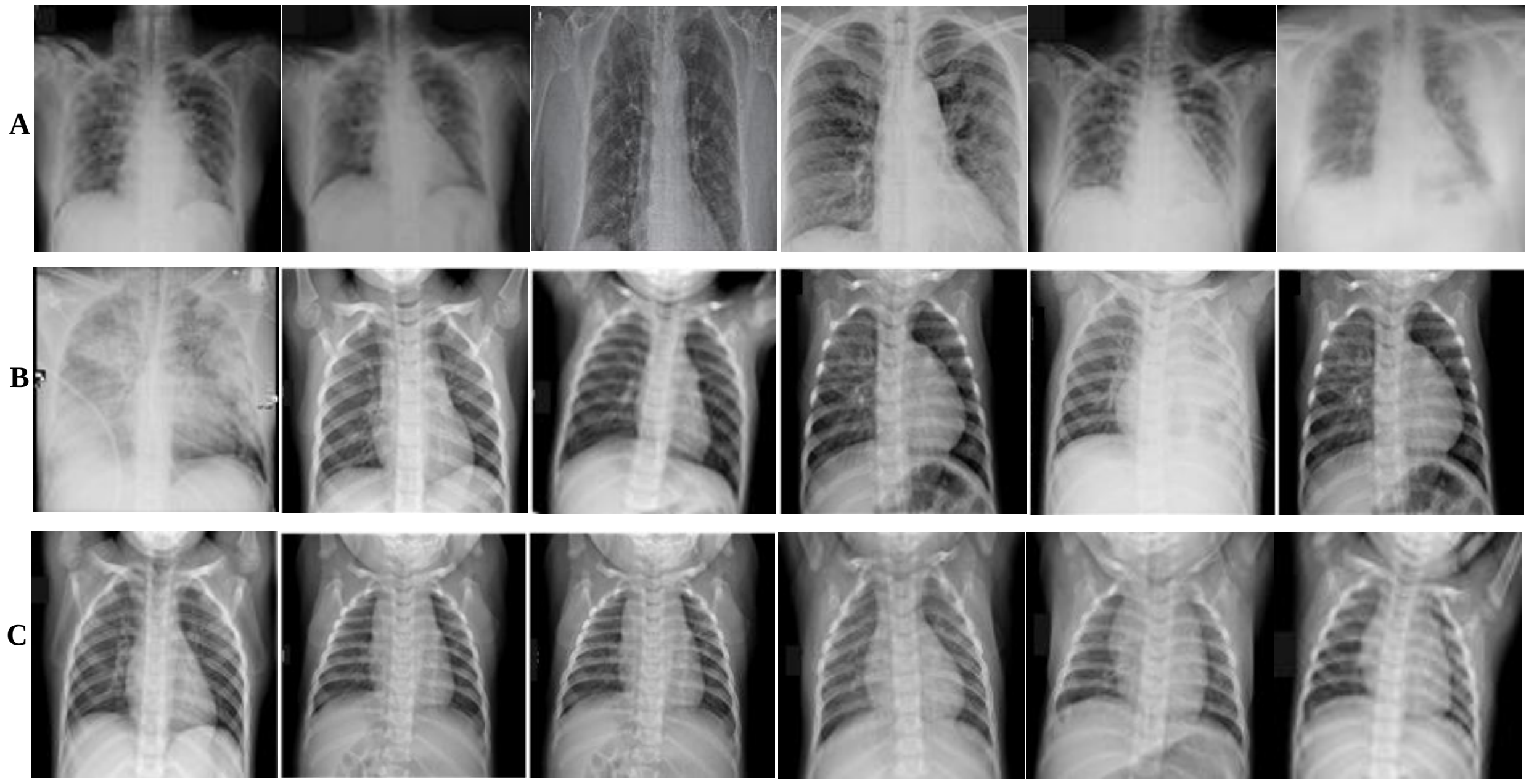

Figure 1: Panel (A), (B), and (C) show COVID-19 infected, Non-COVID-19 infected, and Healthy images, respectively.

### 3.1. CoV-Healthy-6k Dataset

For this study, initially COVID-19 vs Healthy individuals' dataset has been built. The COVID-19 X-Ray images used in this research are collected from [30]. The Healthy individuals' dataset is obtained both from [30] and Kaggle repository [31]. The accessed repositories contain images from multiple publicly accessible sources, and hospitals, and these X-Rays images are verified from the radiologists. The new dataset consists of 3224 images from both COVID-19 infected and Healthy individuals. The advantage of using an open-source dataset is that it can easily implement other different DL models to evaluate the effectiveness of the proposed CB-STM-RENet technique.

### 3.2. CoV-NonCoV-10k Dataset

This dataset consisted of COVID-19 infected and non-COVID-19 chest X-Ray images. The non-COVID-19 X-Rays includes both Healthy and non-COVID-19 infected individuals. These X-Ray images are collected from [30], whereas the same set of Healthy samples are also used as defined in CoV-Healthy-6k Dataset. In non-COVID-19 samples, the disease is caused by different viral and bacterial infections other than COVID-19. This dataset contains total 9538 images, out of which both the COVID-19 and non-COVID-19 class includes the 4769 images.

### 3.3. CoV-NonCoV-15k Dataset

We also build a stringent dataset to evaluate the robustness of the proposed technique. For this, CoV-NonCoV-10k dataset is augmented by including additional samples from [27]. This new CoV-NonCoV-15k dataset is imbalanced and consisted of 15127 total images; out of which 5223 and 9904 images are from COVID-19 infected and non-COVID-19 individuals, respectively.

## 4. Deep Channel Boosting based CNN for COVID-19 Detection

This work proposes a new technique based on CB-CNN for automated detection of COVID-19 in chest X-Ray images. The proposed technique targets the discrimination of COVID-19 infected from both non-COVID-19 infected and Healthy individual. In this regard, a new CNN classifier based on novel split-transform-merge (STM) block [32] is developed that systematically implements RE-based operations for the learning of COVID-19 specific patterns and terms as “STM-RENet”. This architecture is also known as “PIEAS Classification Network-4 (PC Net-4)”. The learning capacity of the proposed CNN is enhanced using Channel Boosting to improve the detection rate while maintaining high precision. The CB-CNN is termed as “CB-STM-RENet” or “PIEAS Classification Network-5 (PC Net-5)”. The performance of the proposed technique is compared with several existing CNNs by implementing them from scratch as well as by adapting them using TL on X-Ray dataset for COVID-19 detection. The overall workflow is shown in Figure 2.

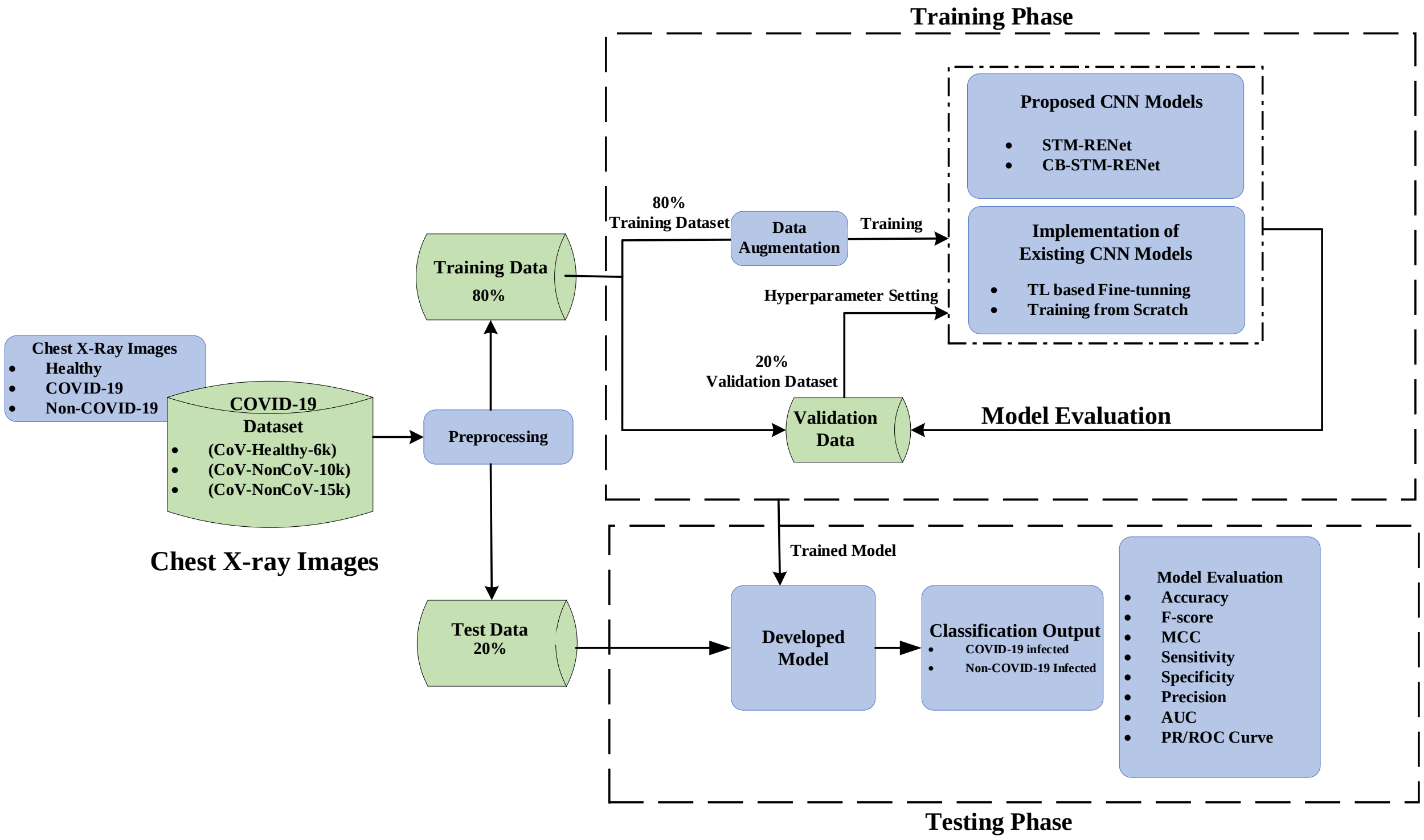

Figure 2: Overview of the workflow for the proposed COVID-19 detection framework.

## 4.1. Proposed STM-RENet

Deep CNNs have been rigorously used in image processing applications because of their strong pattern mining ability. CNN exploits the structural information of the image using convolution operation and dynamically extracts feature hierarchies according to the target application. Multiple innovations in the CNN design have raised their use in medical image classification, detection and pattern discovery tasks [33], [34].

In this work, a new COVID-19 pneumonia specific CNN architecture has been proposed based on the novel split-transform-merge block (STM) and RE-based feature extraction proposed in Khan et al. study [35]. This new architecture is named as STM based RENet (STM-RENet) for COVID-19. The architectural design of the new block is illustrated in Figure 3. The proposed block consists of three sub-branches. The concept of RE-based feature extraction is systematically employed at each branch using max and average pooling in combination with convolution and ReLU activation to capture discriminating feature at a high level.

The STM-RENet mine the patterns from X-Ray dataset by splitting the input into three branches learns the region-specific variations and their characteristic boundaries using RE-based operator and finally merges the output from multiple paths using concatenation operation. In STM-RENet, two STM blocks with the same topology are stacked one after another to extract a diverse set of abstract level features. This idea helps the STM-RENet in extracting a diverse set of variations in the input feature maps.

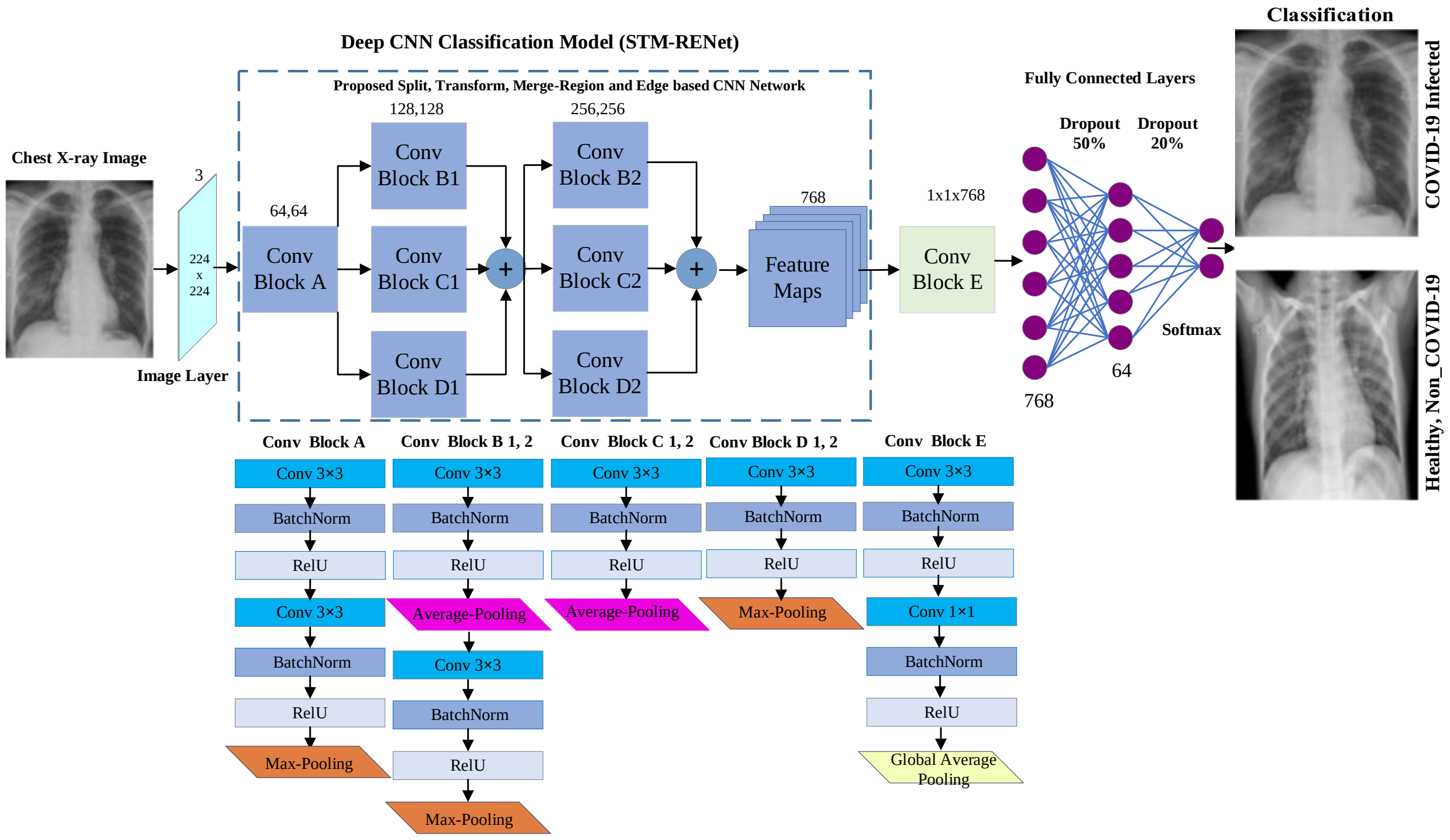

Figure 3: Architectural details of the proposed STM-RENet.

## 4.2. Proposed Channel Boosted STM-RENet (CB-STM-RENet)

Radiographic data exhibits large variations in images and thus, a robust CNN model is required for good discrimination. The discrimination ability of the proposed STM-RENet is enhanced by exploiting Channel Boosting. The idea of Channel Boosting is proposed by Khan et al. [36], [37] for solving complex problems. In the proposed technique, Channel Boosting is performed by generating auxiliary feature channels from two pre-trained networks via TL to improve the performance of STM-RENet.

### 4.2.1. Significance of using Transfer Learning (TL)

TL is a type of machine learning, which allows leveraging the knowledge of existing techniques for new tasks. TL can be exploited in different ways for multiple tasks, but the most often employed approaches for knowledge utilization are 1) instance-based TL, 2) feature-space based TL, 3) parameter exploitation based TL, and 4) Relation-knowledge based TL [38], [39].

Feature space-based TL is often used for solving image classification and pattern recognition related tasks. For this, pre-trained architecture is adapted to the target domain by fine-tuning network layers or by adding additional layers according to the target domain task. This is also commonly known as domain adaptation. Supervised domain adaptation-based TL using pre-trained deep CNNs has been substantially adopted for solving medical imaging tasks. This can help in providing a useful set of feature descriptors that are learnt from the source domain to effectively apply in a target domain by adapting them to target task via fine-tuning. This reduces the calibration efforts (hyper-parameter selection) that are particularly difficult in deep CNNs because of the vast number of hyperparameters and considerable training time [40], [41].

#### 4.2.2. Significance of using Auxiliary Channels

CNNs with varied architectural designs have different feature learning capacities. Multiple channels learnt from different deep CNNs exhibit multi-level information. These channels represent different patterns, which may help in precisely explaining class-specific characteristics. Combination of diverse-level abstractions learned from multiple channels may improve both the global and local representation of the image. The concatenation of auxiliary and original channels gives the idea of intelligent feature-space based ensemble; whereby the single learner takes the final decision by analyzing multiple image specific patterns [42].

#### 4.2.3. Proposed Channel Boosted Architectural Design

In this work, we utilized supervised domain adaptation-based TL by exploiting two different pre-trained deep CNNs. These deep CNNs vary in architectural design that enables each model to learn different feature descriptors and encapsulate diverse level of radiological information from chest X-Ray. These two fine-tuned deep CNNs are termed as auxiliary learner 1 and 2. The purpose of Channel Boosting is to improve the discriminative capability of the proposed CB-STM-RENet model. The architectural details of CB-STM-RENet are illustrated in Figure 4.

$$C_{b}=J_k\left(C_i, \left[R_1, \dots, R_j\right], \left[V_1, \dots, V_j\right]\right) \tag{1}$$

$$F_l^k = F_c(C_b, f_l) \tag{2}$$

In Eq. (1), $C_i$ shows the STM-RENet original channels, whereas $R_j$ $and$ $V_j$ are up to $j^{th}$ number of auxiliary channels generated by TL-based based fine-tuned auxiliary learner 1 and 2, respectively. $J_k(.)$ concatenates the original STM-RENet channels with the auxiliary channels to generates the CB input $C_b$ for the classifier. Eq. (2) illustrates the $k^{th}$ resultant feature maps $F_l^k$, which is generated by convolving the boosted input $C_b$ with filter of $f_l$ layer in convolutional block

E. At the end of convolutional block E, global average-pooling is used to minimize the connection intensity. Finally, the fully connected layers are employed to preserve the prominent features, and dropout layers are used to reduce overfitting.

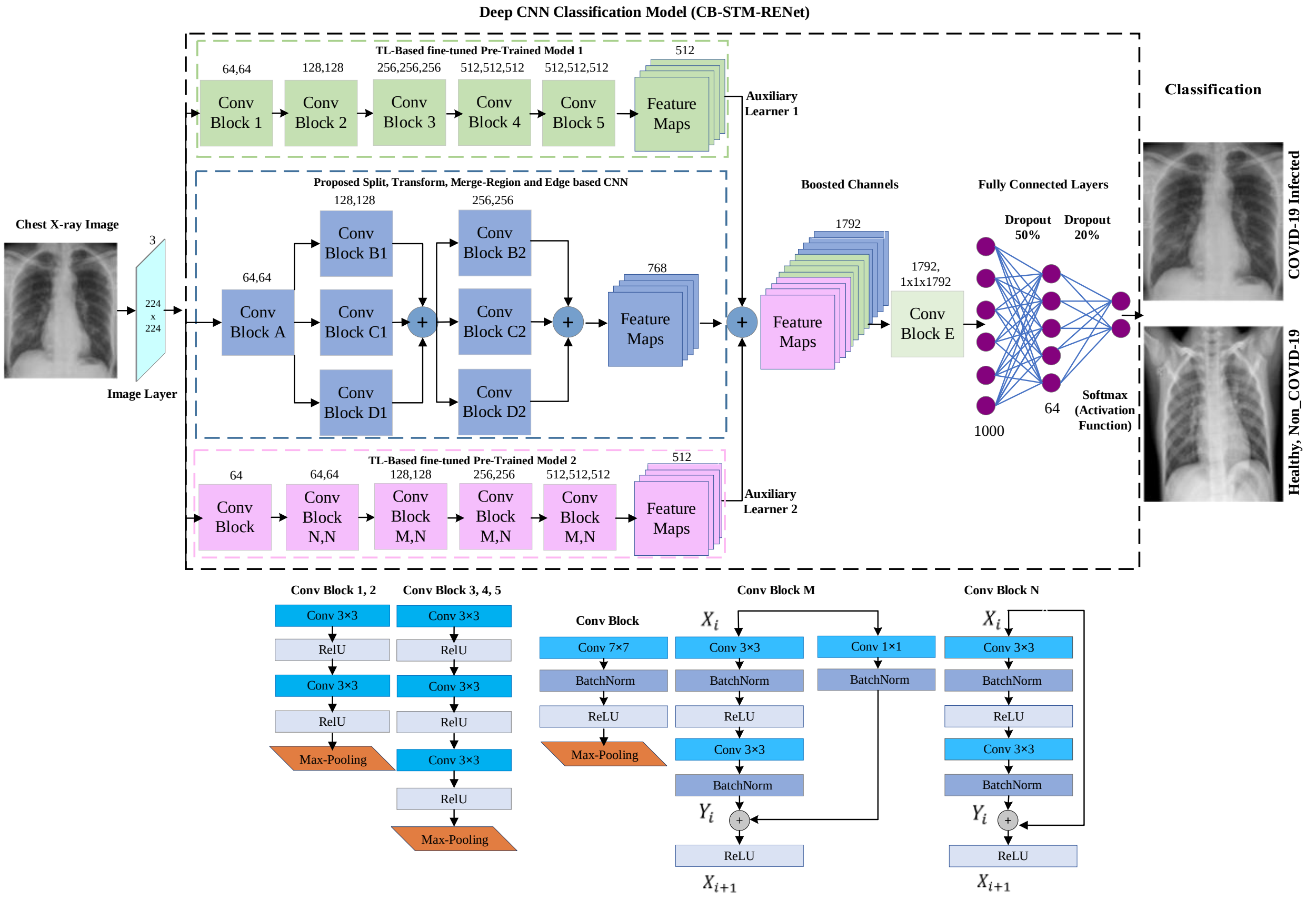

Figure 4: Architectural details of the proposed CB-STM-RENet.

## 4.3. Implementation of the Existing CNNs

For a rigorous assessment of the proposed technique, several existing deep CNNs (Alexnet, VGG-16, VGG-19, Google Net, Inceptionv3, Resnet-18, Resnet-50, Squeeze Net, DenseNet-201, ShuffleNet, Xception) have been implemented and compared with [32], [43]–[49]. These models are initially trained from scratch on X-Ray images for a fair comparison with the proposed technique. The implemented deep models are computationally intensive and require a sufficient amount of training data. Therefore, TL is exploited to train existing deep CNN techniques optimally and to achieve substantial performance on a small amount of data. TL is a type of machine learning, in which models already pre-trained for some task are used for new task by fine-tuning layers of network or by adding some new target specific layers [50], [51]. In this regard, deep CNN models that are pre-trained for ImageNet classification (natural images dataset) are fine-tuned on chest X-Ray images for binary classification.

## 5. Experimental Setup

### 5.1. Dataset Division

Holdout cross-validation scheme is used for the training and evaluation of the deep CNN models. Dataset was divided into train and test set with the ratio of 80:20%. From training dataset, 20% is reserved for model validation and hyperparameter selection. The final evaluation of the model was made on the test set, which was kept separate from training and validation dataset.

### 5.2. Pre-processing

DL models usually overfit on a small size dataset.  Therefore, a sufficient amount of data is required for the effective training and achieving good generalization. Data augmentation is one of the effective ways to improve the generalization of the learning model by incorporating multiple variations in the base dataset. The training samples in this study are augmented by applying different types of transformations, including horizontal and vertical reflections, rotation, and shear. All the images have been resized to 224x224x3 before assigning to CNN for training.

### 5.3. Model Implementation Details

Deep CNN models were trained in an end-to-end method. Stochastic Gradient Descent (SGD) was employed as an optimizer function to reduce cross-entropy loss. Softmax was used for the identification of class probabilities. The training was managed using Piecewise learning rate scheduler by setting an initial value of learning rate as 0.0001 and momentum of 0.95. Some of the CNN Models were trained with a batch size of 16 while others were trained with 32 a batch size for 10 epochs. For each of the CNN model, 95% confidence interval (CI) was computed [52], [53]. The training time for 1 epoch on NVIDIA GeForce GTX Titan X was ~1-2 hours. All the implemented models were trained for all the three different datasets and evaluated on their unseen test sets.

### 5.4. Working Environment

Deep CNN models were built in MATLAB 2019b, and simulations were performed using DL library. All the experimentations were done on a CUDA enabled NVIDIA GeForce GTX Titan X computer, having 64 GB RAM.

## 6. Results

The performance of the proposed technique is evaluated using several performance metrics on an unseen test set and benchmark against well-known existing techniques. Learning plots of the proposed CB-STM-RENet, showing accuracy and loss values for training and validation set is shown in Figure 5. Learning plot suggests that the proposed CB-STM-RENet technique converges to optimal values quickly.

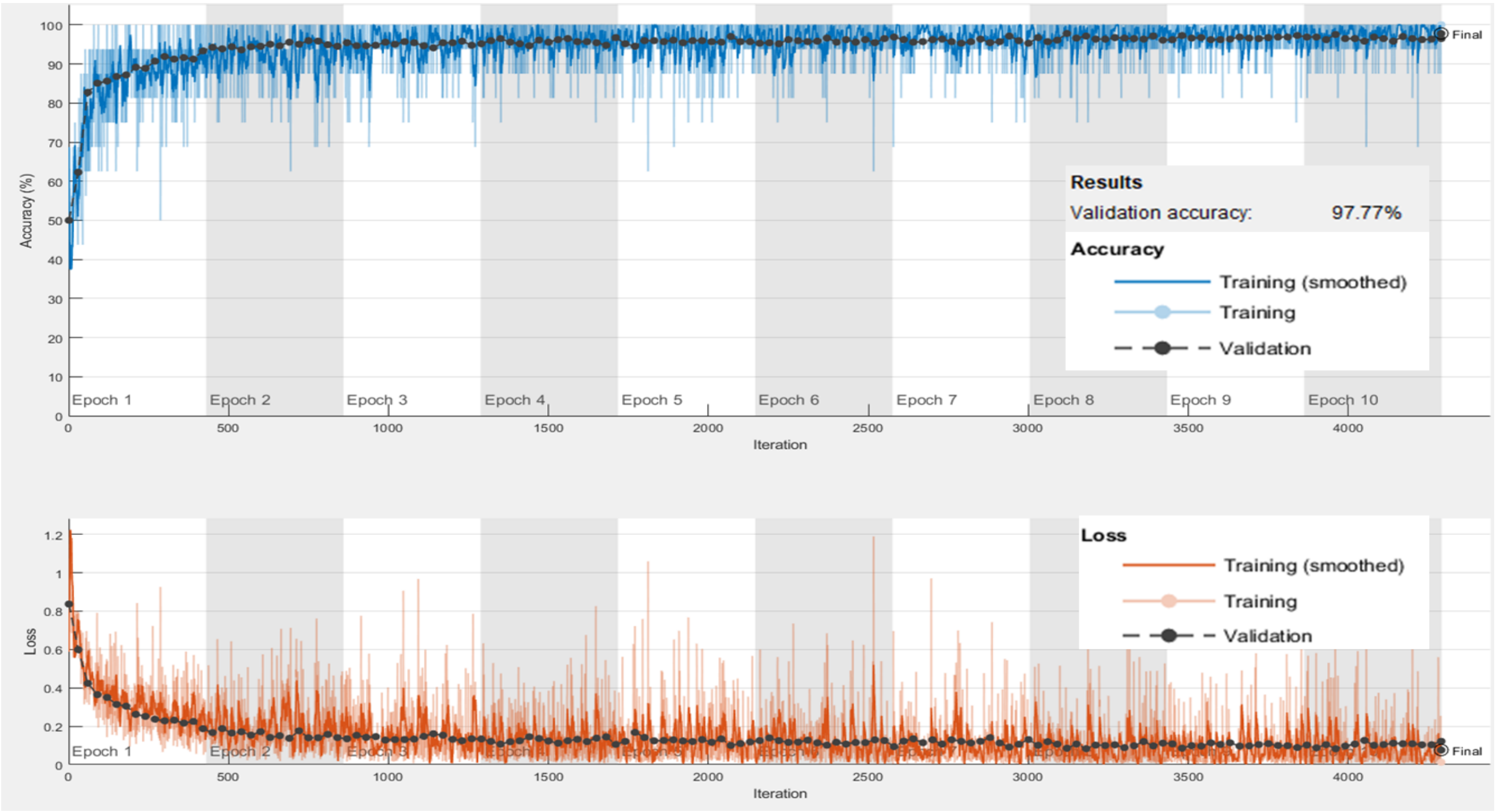

Figure 5: Training Plot of proposed CB-STM-RENet technique on CoV-NonCoV-15k dataset.

### 6.1. Performance Evaluation Metrics

The discrimination ability of the proposed technique is evaluated using accuracy (Acc) and area under the receiver operative characteristic (AUC-ROC) curve for a balanced dataset, whereas F-score and AUC of precision and recall (PR) curve are used as a performance metric for imbalanced dataset. Additionally, Mathew Correlation Coefficient (MCC) is also computed for unbiased estimation as it considers all the examples from COVID-19 positive and negative classes including both True (TP, TN) and False (FN, FP) predictions. COVID-19 negative class includes both Healthy and non-COVID-19 infected individuals. The details of the qualitative measure such as sensitivity (sen), specificity (spe), precision (pre), TP, TN, FN, and FP are also reported. Examples from COVID-19 positive and negative class that are correctly predicted are known as TP and TN, respectively. Similarly, positive, and negative class examples that are misclassified are referred to as FP and FN, respectively. These performance metrics are mathematically expressed in Eq. 3-8.

Accuracy defines the ratio of COVID-19 positive and COVID-19 negative samples that are correctly classified. COVID-19 negative can be Healthy individuals or patients having other viral/bacterial infection. F-score is a measure of accuracy for the imbalanced dataset. Sensitivity and specificity refer to the ratio of COVID-19 positives and negative patients, respectively that are correctly identified. Precision is the proportion of COVID-19 positive predictions that are made actually correct.

$$\mathrm{Acc} = \frac{\text{True COVID}-19\text{ positives (TP)}+\text{True COVID}-19\text{ negatives (TN)}}{\text{Total samples (TP+TN+FP+FN)}} \times 100 \tag{3}$$

$$\mathrm{Sen} = \frac{\text{True COVID}-19\text{ positives (TP)}}{\text{Total COVID}-19\text{ positive Samples (TP+FN)}} \times 100 \tag{4}$$

$$\mathrm{Spe} = \frac{\text{True COVID}-19\text{ negatives (TN)}}{\text{Total COVID}-19\text{ negative Samples (TN+FP)}} \times 100 \tag{5}$$

$$\mathrm{Pre} = \frac{\text{True COVID}-19\text{ positives (TP)}}{\text{True COVID}-19\text{ positives (TP)}+\text{False COVID}-19\text{ positives (FP)}} \times 100 \tag{6}$$

$$\mathrm{F}-\mathrm{Score} = 2 \times \frac{\mathrm{Pre}\times\mathrm{Sen}}{\mathrm{Pre}+\mathrm{Sen}} \tag{7}$$

$$\mathrm{MCC} = \frac{(\mathrm{TN}\times\mathrm{TP})-(\mathrm{FN}\times\mathrm{FP})}{\sqrt{((\mathrm{FN}+\mathrm{FP})(\mathrm{FP}+\mathrm{TP})(\mathrm{FN}+\mathrm{TN})(\mathrm{FP}+\mathrm{TN}))}} \tag{8}$$

### 6.2. Performance Analysis on CoV-Healthy-6k

Classification results of the proposed STM-RENet with and without Channel Boosting on the test set of CoV-Healthy-6k are shown in Table 1. The classification ability in terms of accuracy (STM-RENet: 97.98%, CB-STM-RENet: 98.53%), F-score (STM-RENet: 0.98, CB-STM-RENet: 0.98) and MCC (STM-RENet: 0.96, CB-STM-RENet: 0.97) show that both models can effectively differentiate the COVID-19 infected patients from Healthy individuals.

## 6.3. Performance Analysis on CoV-NonCoV-10k

The proposed technique is accessed for its effectiveness in discriminating COVID-19 infected from non-COVID-19 infected. Therefore, STM-RENet with and without Channel Boosting is trained on CoV-NonCoV-10k with the same set of parameters and evaluated on the test dataset. Table 1 illustrates the classification results. The performance analysis using various evaluation metrics (accuracy: 97.48%, F-score: 0.98, and MCC: 0.95) clearly shows that Channel Boosting improves the discrimination ability of CNN significantly (shown in Table 1).

## 6.4. Performance Analysis on the Stringent CoV-NonCoV-15k

Generalization of the proposed technique is accessed by evaluating the performance on stringent CoV-NonCoV-15k dataset, as shown in Figure 6. This dataset is imbalance and contains a smaller number of COVD-19 positive patients as compared to non-COVID-19 and Healthy individuals both in training and test set. Table 1 shows the detection results. F-score and AUC show good learning potential and strong discrimination ability of our proposed CB-STM-RENet technique as compared to STM-RENet.

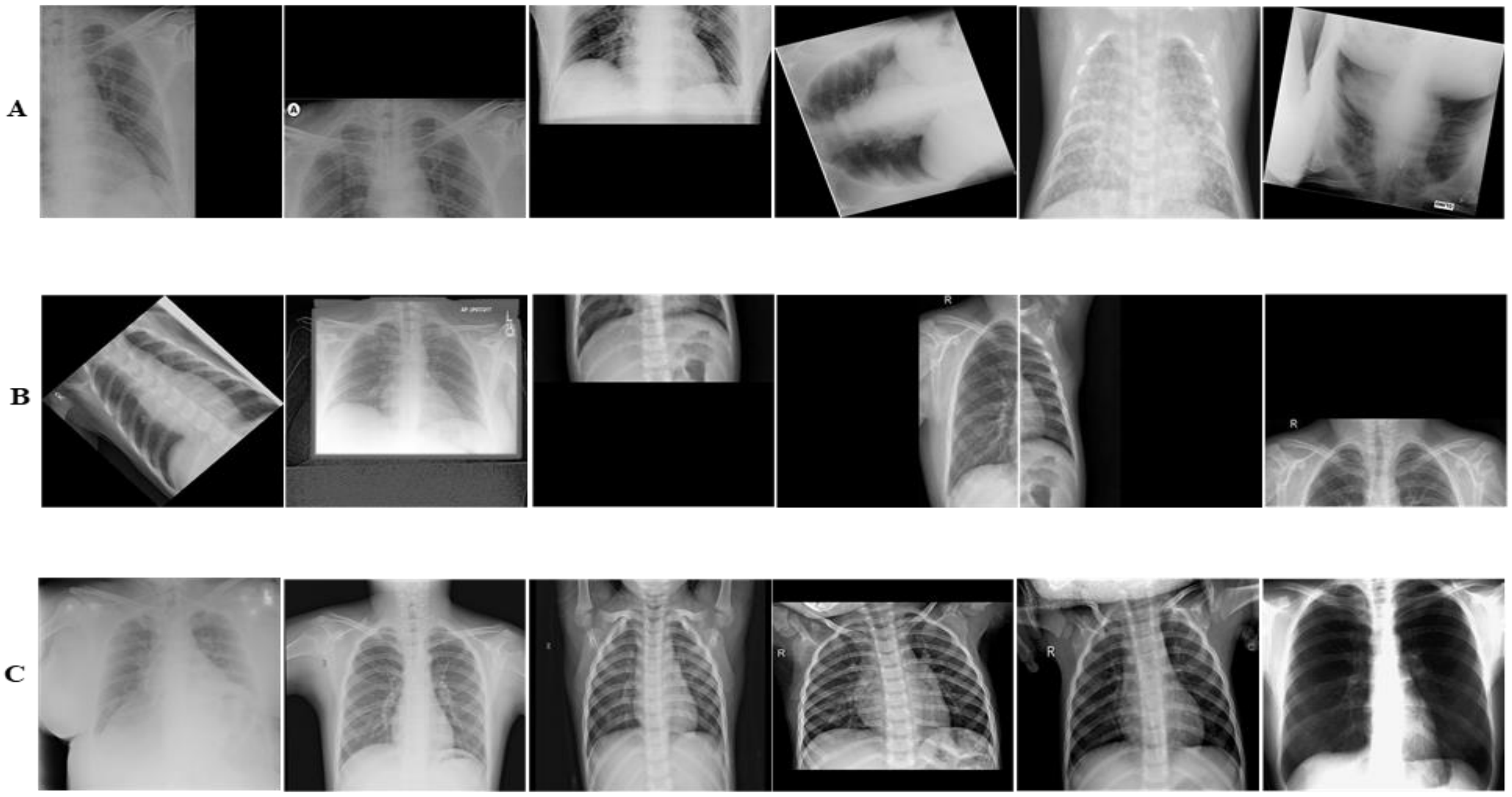

Figure 6: Panel (A), (B), and (C) show COVID-19 infected, Non-COVID-19 infected, and Healthy images, respectively. The images are tough to classify because of having high illumination, translational, rotational, occlusion and missing informational effects.

Table 1: Performance evaluation of the proposed techniques on the test set of CoV-Healthy-6k, CoV-NonCoV-10k and CoV-NonCoV-15k. Standard Error (SE) at 95% CI is reported for sensitivity.

| CoV-Healthy-6k | | | | | | |
|---|---|---|---|---|---|---|
| **Techniques** | **Accuracy %** | **MCC** | **F-score** | **Sensitivity ± SE** | **Precision** | **Specificity** |
| **STM-RE-CBNet** | **98.53** | **0.97** | **0.98** | **0.99± 0.02** | **0.98** | **0.98** |
| **STM-RENet** | 97.98 | 0.96 | 0.98 | 0.97± 0.04 | 0.99 | 0.99 |
| **CoV-NonCoV-10k** | | | | | | |
| **STM-RE-CBNet** | **97.48** | **0.95** | **0.98** | **0.99 ± 0.02** | **0.98** | 0.96 |
| **STM-RENet** | 91.82 | 0.84 | 0.92 | 0.97 ± 0.04 | 0.88 | 0.86 |
| **CoV-NonCoV-15k** | | | | | | |
| **STM-RE-CBNet** | **96.53** | **0.93** | **0.95** | **0.97 ± 0.04** | **0.93** | **0.96** |
| **STM-RENet** | 92.04 | 0.85 | 0.90 | 0.96 ± 0.05 | 0.84 | 0.90 |

Table 2: Performance analysis of the best performing existing CNN techniques on the test set of CoV-Healthy-6k, CoV-NonCoV-10k and CoV-NonCoV-15k. SE at 95% CI is reported for sensitivity.

| CoV-Healthy-6k | | | | | | |
|---|---|---|---|---|---|---|
| **Techniques** | **Accuracy %** | **MCC** | **F-score** | **Sensitivity ± SE** | **Precision** | **Specificity** |
| **Resnet18** | 96.59 | 0.93 | 0.96 | 0.98 ± 0.03 | 0.95 | 0.95 |
| **VGG_16** | 95.74 | 0.91 | 0.95 | 0.96 ± 0.05 | 0.95 | 0.95 |
| **CoV-NonCoV-10k** | | | | | | |
| **ResNet-18** | **90.36** | **0.81** | **0.90** | **0.90 ± 0.12** | **0.90** | **0.91** |
| **VGG-16** | 86.32 | 0.73 | 0.86 | 0.86 ± 0.16 | 0.86 | 0.87 |
| **CoV-NonCoV-15k** | | | | | | |
| **ResNet-18** | **91.20** | **0.84** | **0.88** | **0.95 ± 0.06** | **0.82** | **0.89** |
| **VGG-16** | 88.26 | 0.77 | 0.84 | 0.89 ± 0.13 | 0.80 | 0.88 |

## 6.5. Comparative Analysis with the Existing CNNs

The significance of the proposed architecture and the impact of Channel Boosting is explored by implementing existing deep CNN techniques. Existing techniques with different CNN architectural blocks are implemented from scratch as well as well fine-tuned using TL. Figure 7 and Table 2 shows the result of the best performing techniques on a test set of CoV-Healthy-6k, CoV-NonCoV-10k and CoV-NonCoV-15k. Table 3 shows the results of all of these existing techniques on CoV-Healthy-6k. The evaluation metrics suggest that TL improves the learning of discriminating patterns for COVID-19 classification.

The performance comparison based on accuracy, F-score and MCC suggest that the proposed technique STM-RENet with and without Channel Boosting outperformed the existing techniques (Table 1-3). The performance gain of the proposed CB-STM-RENet as compared to the highest performing existing CNN technique (ResNet) is illustrated in Figure 7.

Table 3: Performance of existing CNN techniques on the test set of CoV-Healthy-6k. SE at 95% CI is reported for sensitivity.

| Techniques | Trained from scratch and TL-based fine-tuned CNNs | | | | | |
|---|---|---|---|---|---|---|
| | Accuracy % | MCC | F-score | Sensitivity ± SE | Precision | Specificity |
| **ShuffleNet** | 84.88 | 0.75 | 0.86 | 0.95 ± 0.06 | 0.79 | 0.75 |
| **TL_ShuffleNet** | 96.59 | 0.93 | 0.96 | 0.96 ± 0.05 | 0.97 | 0.97 |
| **Inceptionv3** | 93.80 | 0.86 | 0.93 | 0.95 ± 0.06 | 0.92 | 0.92 |
| **TL_Inceptionv3** | 96.51 | 0.93 | 0.96 | 0.97 ± 0.04 | 0.96 | 0.96 |
| **Alexnet** | 94.50 | 0.89 | 0.94 | 0.92 ± 0.09 | 0.97 | 0.97 |
| **TL_Alexnet** | 95.74 | 0.91 | 0.95 | 0.96 ± 0.05 | 0.95 | 0.95 |
| **DenseNet201** | 95.50 | 0.91 | 0.95 | 0.95 ± 0.06 | 0.96 | 0.96 |
| **TL_DenseNet201** | 96.51 | 0.93 | 0.96 | 0.96 ± 0.05 | 0.97 | 0.97 |
| **Xception** | 95.74 | 0.91 | 0.95 | 0.94 ± 0.07 | 0.97 | 0.97 |
| **TL_Xception** | 96.43 | 0.93 | 0.96 | 0.96 ± 0.05 | 0.97 | 0.97 |
| **TL_ VGG_16** | 97.05 | 0.94 | 0.97 | 0.97 ± 0.04 | 0.97 | 0.97 |
| **Resnet50** | 96.28 | 0.92 | 0.96 | 0.95 ± 0.06 | 0.98 | 0.98 |
| **TL_Resnet50** | 97.05 | 0.94 | 0.97 | 0.97 ± 0.04 | 0.97 | 0.98 |
| **TL_Resnet18** | 97.13 | 0.94 | 0.97 | 0.97 ± 0.04 | 0.98 | 0.98 |

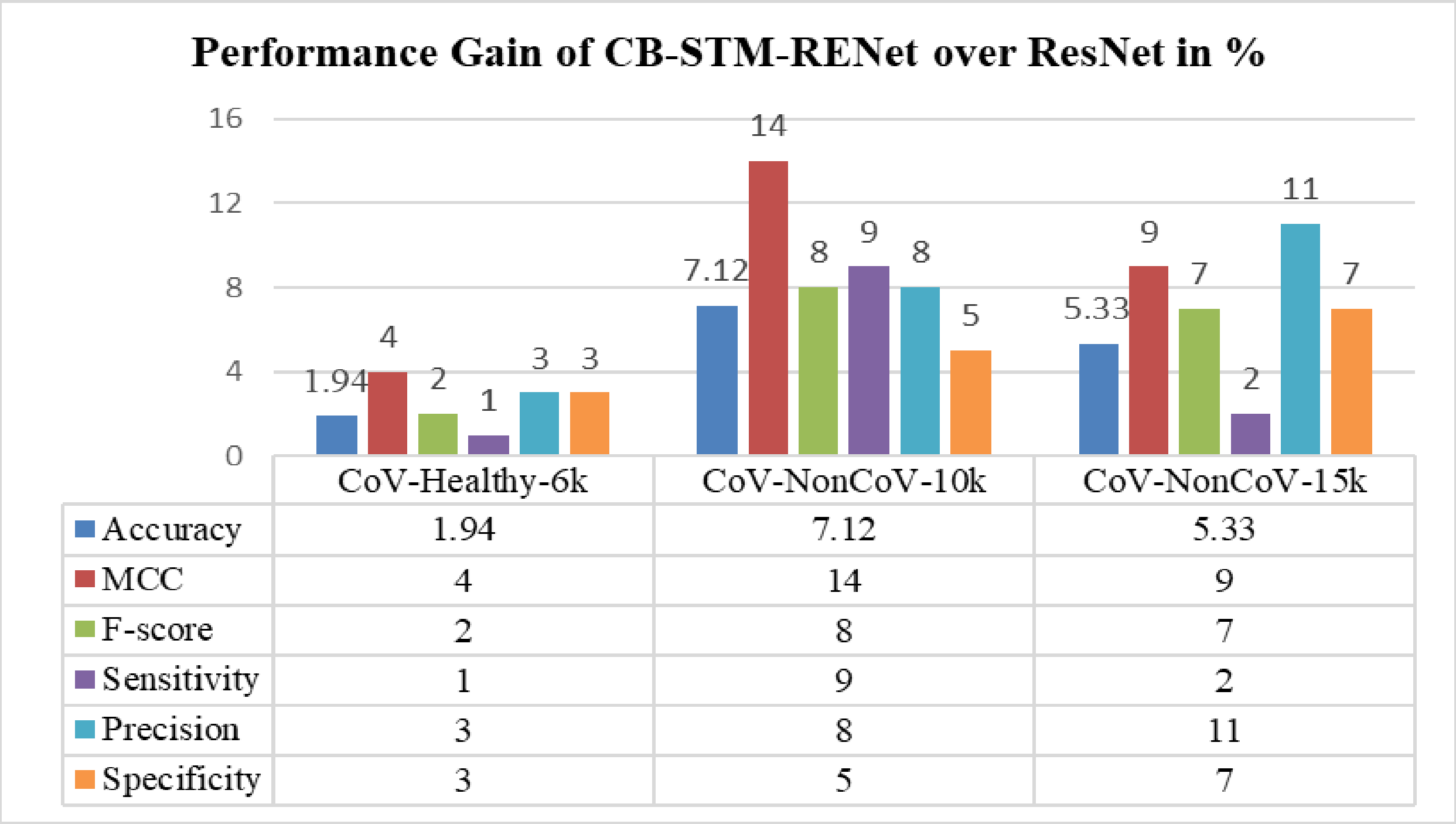

Figure 7: Performance gain of the proposed CB-STM-RENet techniques as compared with the ResNet.

## 6.6. Feature Space Visualization

Feature space learnt by the proposed CB-STM-RENet technique and ResNet is analyzed to understand the decision-making behavior better. Figure 8 shows the projection of the first two principal components of feature space learned by the proposed CB-STM-RENet and ResNet for the test dataset. It is evident from 2D plots that the proposed CB-STM-RENet shows the highest discriminative capability (segregation of COVID-19 positive from Non-COVID-19) as compared with ResNet on the test dataset of CoV-Healthy-6k, CoV-NonCoV-10k and CoV-NonCoV-15k, respectively.

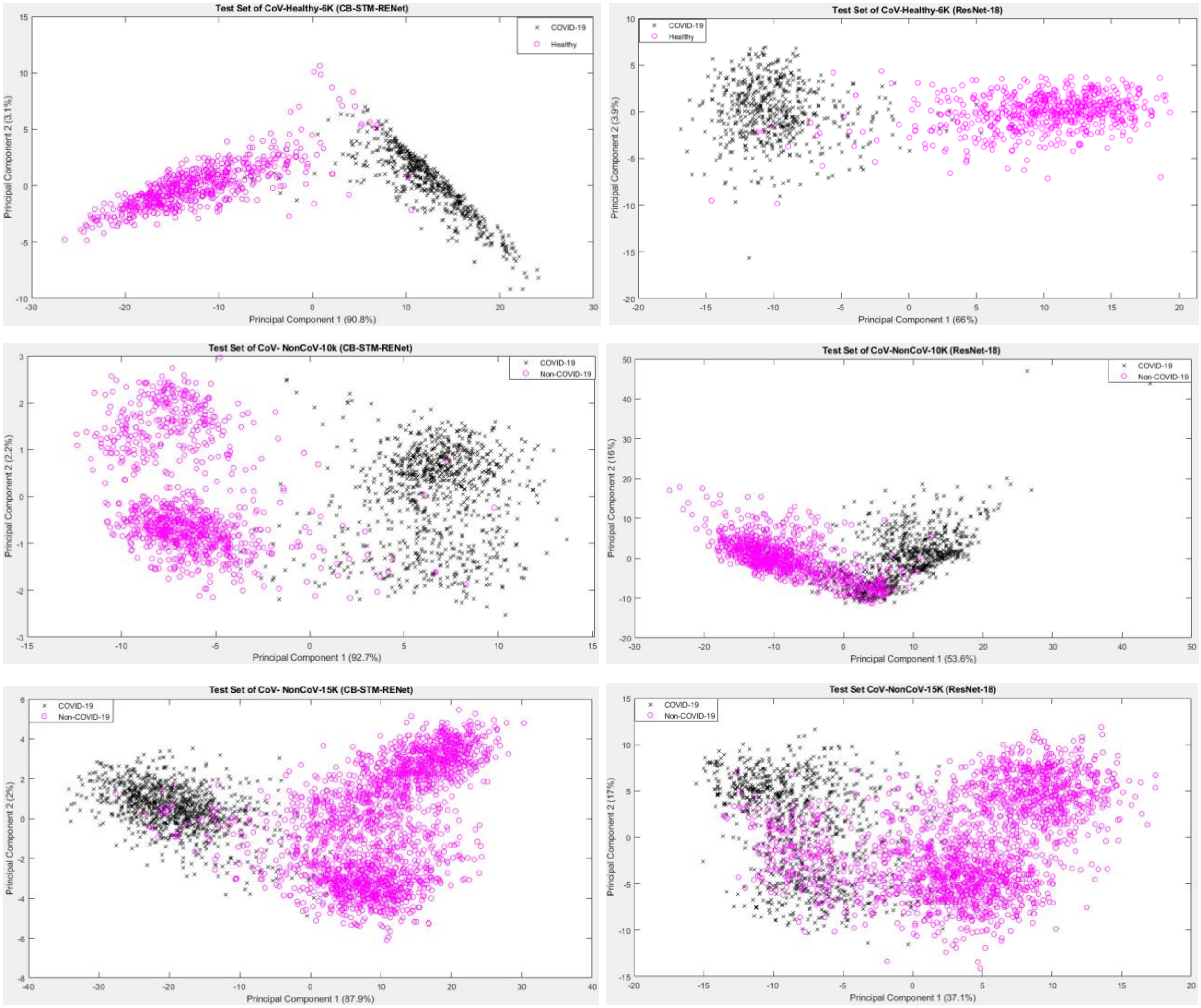

Figure 8: Feature visualization of the proposed CB-STM-RENet and ResNet on test set dataset.

### 6.7. Detection Rate Analysis

Significant detection rate is needed in COVID-19 diagnostic system for limiting infection spread and patient treatment. Therefore, the detection rate (number of correctly identified COVID-19 positive patients) is explored along with the precision of the proposed technique for all three test sets. The detection rate and precision for the proposed CB-STM-RENet and best performing existing techniques are reported in Figure 9 and Table 1-2.

The quantitative statistics exhibit that the proposed technique with and without Channel Boosting achieved the highest detection rate (ranging from 96-99%) with the minimum number of False positives. CB-STM-RENet significantly decreases the number of False negatives and positives, as shown in Figure 9. The substantial precision suggests that our proposed technique with Channel Boosting significantly reduced the miss-detection rate (ranging from 1-7%) and is able to screen the individuals precisely. High precision means very few Healthy individuals or non-COVID-19 patients will be Falsely diagnosed with COVID-19 infection and thus result in lessening the burden on radiologists.

## 6.8. Evaluation of Diagnostic Ability of the Proposed Technique

ROC and PR curve have a significant role in accessing the optimal diagnostic cutoff for the classifier. These curves graphically illustrate the discrimination ability of the classifier at a whole range of possible values [54]. Figure 10 shows ROC curves for the proposed and existing techniques for CoV-Healthy-6k and CoV-NonCoV-10k datasets, whereas both the curves, ROC and PR are reported for CoV-NonCoV-15k dataset because of its imbalance nature.

It is evident from ROC and PR based quantitative analysis that the proposed technique, both with and without Channel Boosting at different cutoffs, shows significant diagnostic accuracy. Figure 10 shows that our proposed technique with Channel Boosting achieved an AUC-ROC of 0.99 on both the datasets (CoV-Healthy-6k and CoV-NonCoV-10k) and 0.98 AUC-PR for CoV-NonCoV-15k. The high value of AUC recommends that the proposed technique with Channel Boosting upholds high sensitivity with low False detection rate and performs well as a whole for COVID-19 patients' screening.

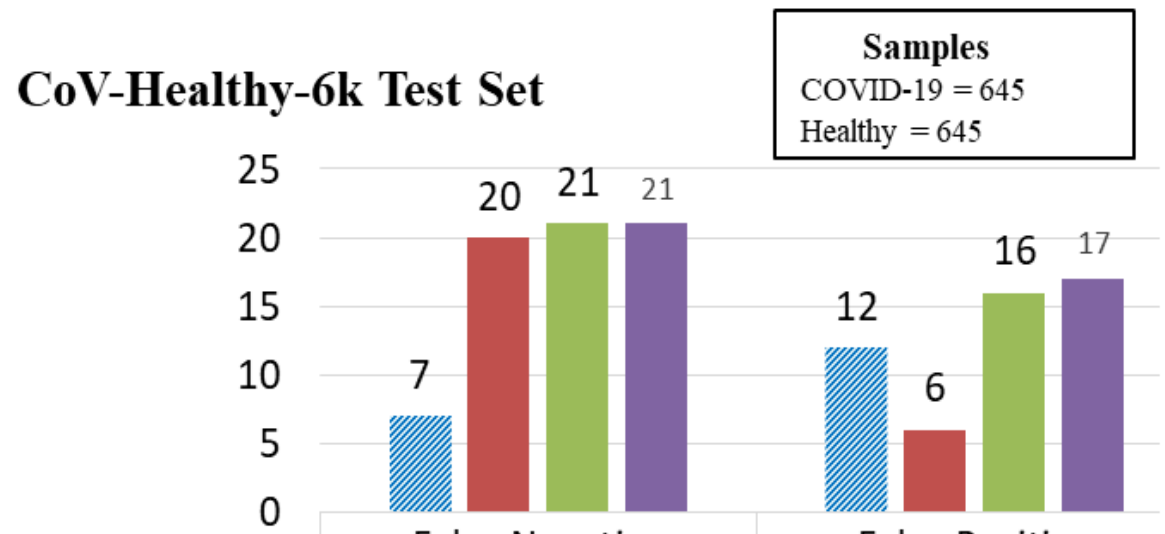

| | False Negative | False Positive |
|---|---|---|
| CB-STM-RENet | 7 | 12 |
| STM-RENet | 20 | 6 |
| ResNet-18 | 21 | 16 |
| VGG-16 | 21 | 17 |

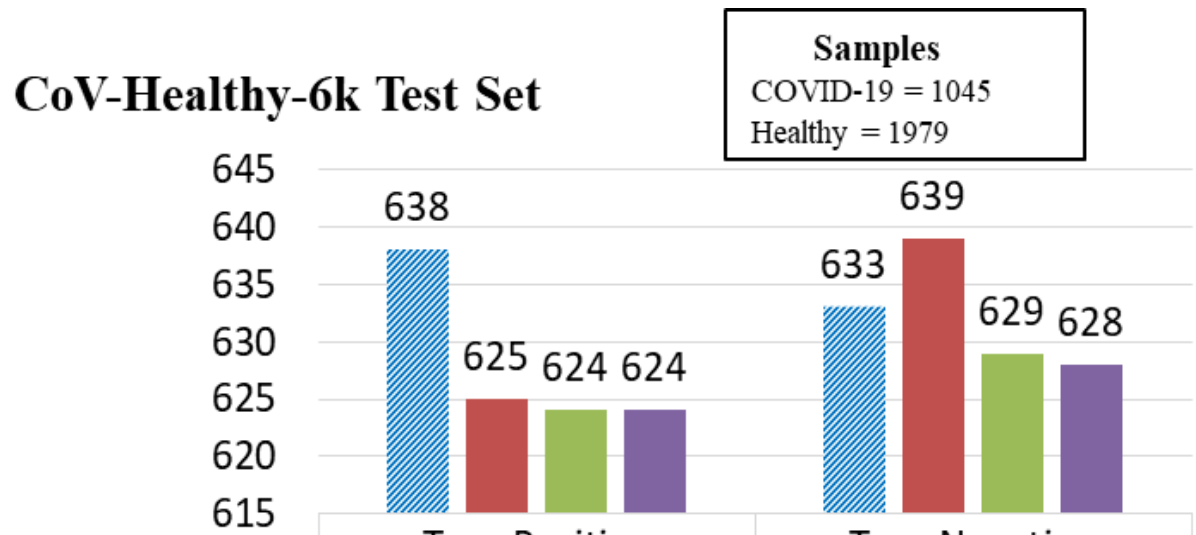

| | True Positive | True Negative |
|---|---|---|
| CB-STM-RENet | 638 | 633 |
| STM-RENet | 625 | 639 |
| ResNet-18 | 624 | 629 |
| VGG-16 | 624 | 628 |

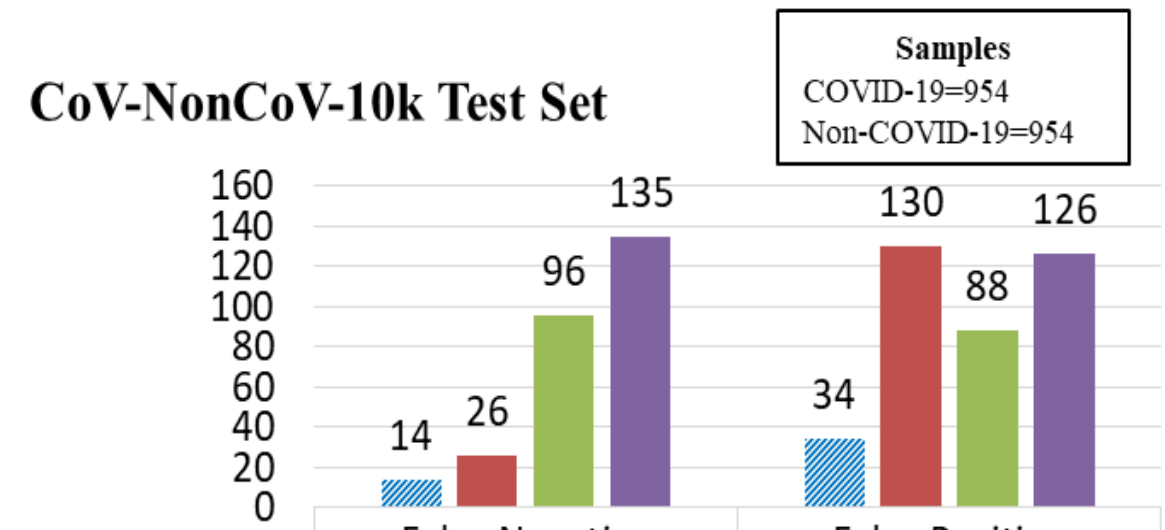

| | False Negative | False Positive |
|---|---|---|
| CB-STM-RENet | 14 | 34 |
| STM-RENet | 26 | 130 |
| ResNet-18 | 96 | 88 |
| VGG-16 | 135 | 126 |

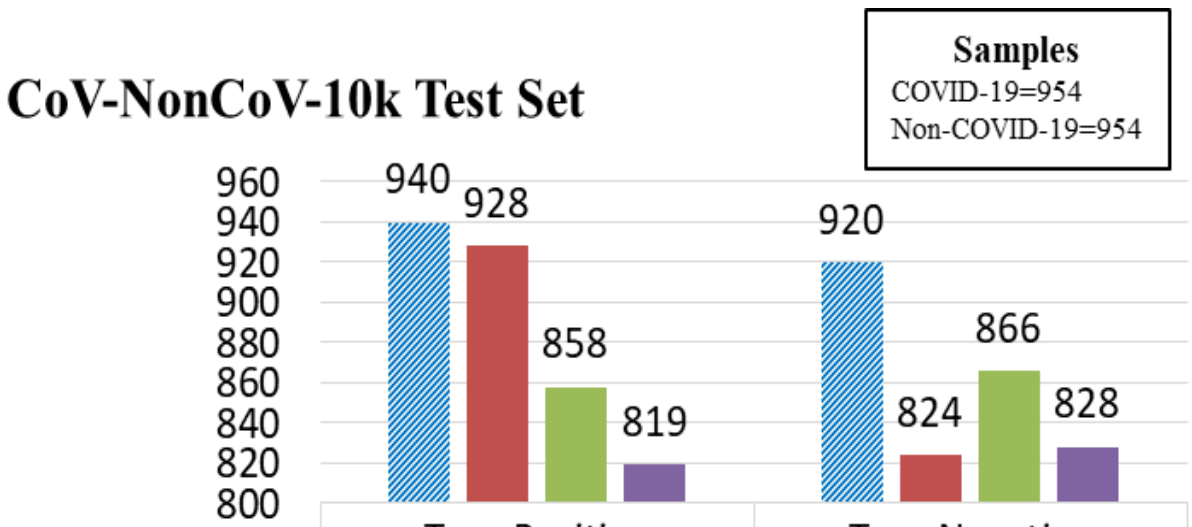

| | True Positive | True Negative |
|---|---|---|
| CB-STM-RENet | 940 | 920 |
| STM-RENet | 928 | 824 |
| ResNet-18 | 858 | 866 |
| VGG-16 | 819 | 828 |

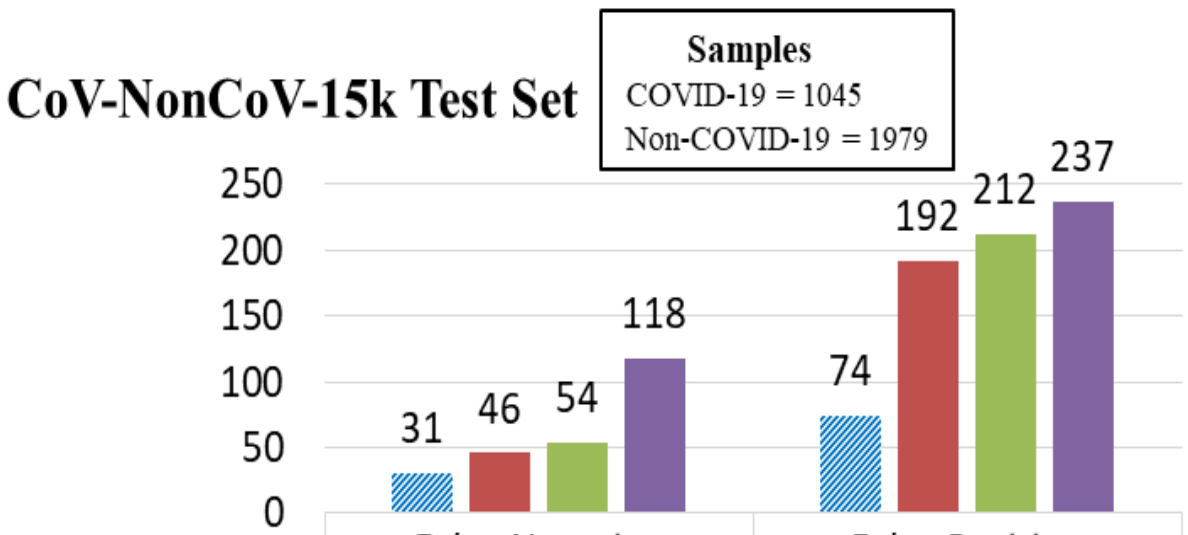

| | False Negative | False Positive |
|---|---|---|
| CB-STM-RENet | 31 | 74 |
| STM-RENet | 46 | 192 |
| ResNet-18 | 54 | 212 |
| VGG-16 | 118 | 237 |

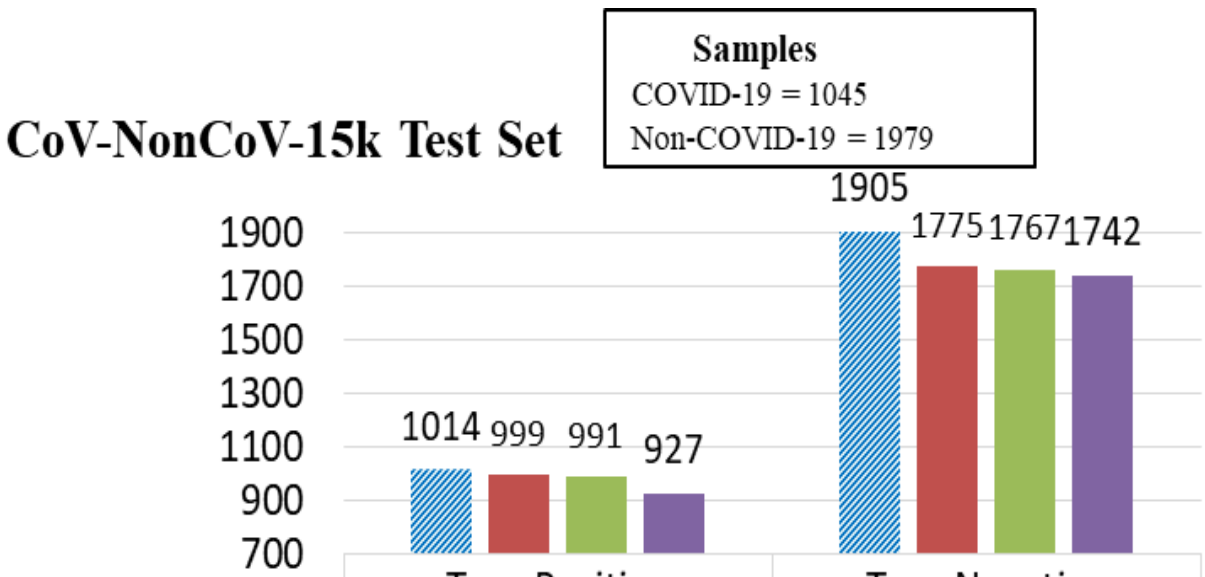

| | True Positive | True Negative |
|---|---|---|
| CB-STM-RENet | 1014 | 1905 |
| STM-RENet | 999 | 1775 |
| ResNet-18 | 991 | 1767 |
| VGG-16 | 927 | 1742 |

Figure 9: Detection rate analysis of the proposed CB-STM-RENet with the existing CNN techniques.

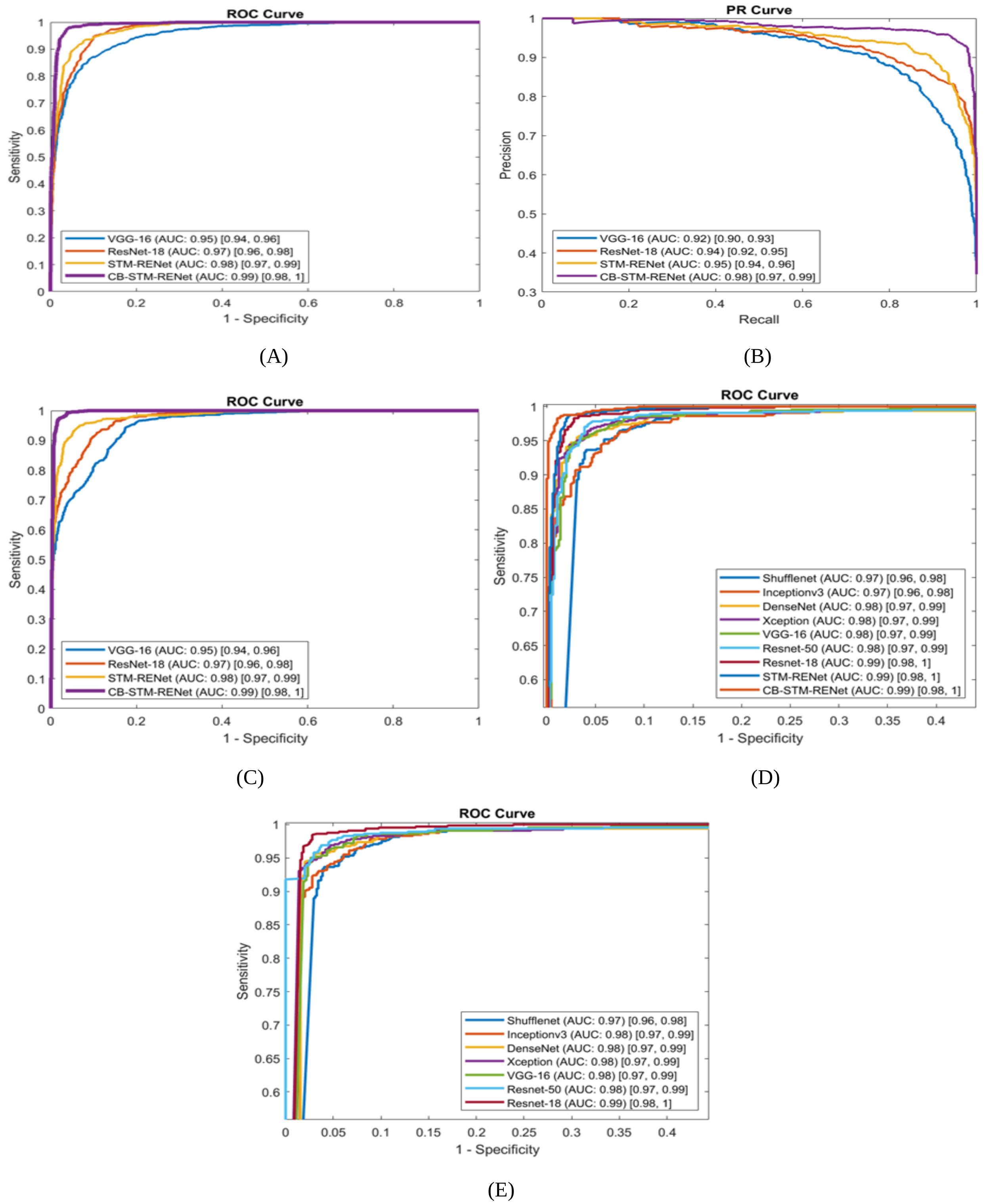

Figure 10: ROC and PR Curve for the proposed STM-RENet, CB-STM-RENet, and existing CNN techniques. (A) & (B) are reported for CoV-NonCOV-15k, (C) for CoV-NonCOV-10k, while (D) & (E) for CoV-Healthy-6k. AUC CI at 95% is shown in square bracket.

## 7. Conclusion

COVID-19 is a contagious viral infectious disease that has affected a broad spectrum of the global population. The high transmissibility and pathogenic nature make the early detection of infected individuals vital for commendably fighting COVID-19. Radiographic examination of chest X-Rays is one of the fastest and key screening approaches. Chest X-Ray images exhibit characteristic patterns that are associated with COIVD-19 abnormalities. In this work, a new deep Channel Boosting based CNN is proposed to detect COVID-19 in chest X-Rays by discerning radiographic patterns. The proposed technique is benchmarked by implementing and comparing with several existing CNN architectures and techniques reported in the literature. The performance comparison of the proposed deep CB-STM-RENet with the existing techniques exhibits high classification performance, both in discriminating COVID-19 chest infections from Healthy, as well as, other types of chest infections. CB-STM-RENet provides highest performance on all the three datasets, especially on stringent CoV-NonCOV-15k dataset (Acc:96.53%, F-score:0.95, MCC:0.93, and AUC:0.98). The good detection rate (97%), high precision (93%), and low False positives of the proposed technique suggest that it can be adapted for the diagnosis of COVID-19 infection.

### Acknowledgements

This work was conducted with the support of PIEAS IT endowment fund and HEC indigenous Scholarship under the Pakistan Higher Education Commission (HEC). We would like to thank Abdur Rehman, Ayesha Salar and Aleena Ijaz from Pattern Recognition Lab (PR-Lab), PIEAS for providing helping material.

### Conflicts of interest

Authors declared no conflict of interest.

### Code availability

The test code is available at https://github.com/PRLAB21/COVID-19-Detection-System-using-Chest-X-Ray-Images.